\documentclass[aps,prb,twocolumn,showpacs,groupedaddress]{revtex4}

\usepackage{graphicx}

\begin{document}


\title{Electron-electron interaction corrections to the thermal conductivity
in disordered conductors }


\author{Douglas R Niven}
\author{Robert A Smith}
\affiliation{School of Physics and Astronomy, University of
Birmingham, Edgbaston, Birmingham B15 2TT, ENGLAND}


\date{\today}

\begin{abstract}
We evaluate the electron-electron interaction corrections to the electronic
thermal conductivity in a disordered conductor in the diffusive regime. We
use a diagrammatic many-body method  analogous to that of Altshuler and Aronov 
for the electrical conductivity. We derive results in one, two and three
dimensions for both the singlet and triplet channels, and in all cases find
that the Wiedemann-Franz law is violated. 
\end{abstract}


\maketitle

\section{Introduction}

The effect of electron-electron interaction on the electrical
conductivity of disordered systems has been extensively investigated both
theoretically and experimentally over the past two 
decades\cite{AA85,Zala01}. 
There are two
main types of correction to the Drude electrical conductivity which can have
similar magnitudes and temperature dependences: weak localization and
interaction effects. Weak localization is due to interference between
pairs of time-reversed scattering trajectories of electrons from impurities;
this effect can occur in a non-interacting system. Interaction effects are
due to the increased effective electron-electron interaction strength due
to incomplete screening by diffusively moving electrons. Experimentally the
two effects can be distinguished by applying a magnetic field; weak
localization is suppressed whilst interaction effects are not.

In comparison there has been relatively little work done on the thermal
conductivity, either theoretically or experimentally. This is largely because
thermal conductivity is hard to measure in low-dimensional systems at low
temperature, and it is difficult to separate the electronic and lattice
contributions. It is therefore doubly hard to observe the disorder-driven
corrections to the electronic thermal conductivity. Moreover there have been
theoretical predictions\cite{Ches60,Cast88} 
that the Wiedemann-Franz law holds, which allows one
to deduce the thermal conductivity, $\kappa$, directly from the electrical 
conductivity, $\sigma$,
\begin{equation}
\label{WF}
{\kappa\over\sigma}={\pi^2\over 3}\left({k_B\over e}\right)^2T=L_0T,
\end{equation}
where $k_B$ is Boltzmann's constant, $e$ is the electronic charge, 
$T$ is the temperature, and $L_0$
is known as the Lorenz number. Clearly it is important to establish whether
the Wiedemann-Franz law is valid in interacting disordered systems; if it is,
there is nothing to be gained from measuring the thermal conductivity in
addition to the electrical conductivity.

In this paper we evaluate the interaction corrections to the thermal conductivity
in one, two and three dimensions in both the singlet and triplet channels.
We find that the Wiedemann-Franz law is violated in all cases. 
Our results are presented in detail in Table 1. We have not included a result
for the singlet channel in one dimension since this is just one third of the
corresponding triplet channel result with $F$ replaced by an effective singlet
channel interaction $F_s=\kappa_3^2a^2\ln{(D\kappa_3^2/T)}$, as explained in
the figure caption. Each term consists
of two pieces: the first piece leads to violation of the Wiedemann-Franz law;
the second piece does not. The exception is the triplet channel in two 
dimensions which does not possess a violating piece. Note that in each case
both terms have the same parametric dependences; it is only the constant
prefactors which are different.

Our calculation
is the exact analogue of the original Altshuler-Aronov calculation for
electrical conductivity\cite{AA79}. 
There have been several previous attempts at this
calculation\cite{Cast88,Liv91,Arfi92}; 
however these are in disagreement with each other, and we believe
all of them to be incorrect. The fact that it has taken so long (over twenty
years) to get the correct result for the thermal conductivity is due to three
inherent difficulties in the problem. First, the heat-current operator is not
uniquely defined, and some choices of definition will be renormalized by the
electron-electron interaction\cite{Liv91,Lang62}. 
Second, the heat-current operator has opposite
electron-hole parity to the charge-current operator, which leads to the presence
of an extra ``heat-drag'' diagram for the thermal conductivity which is
vanishingly small for the electrical conductivity. (We use the term 
``heat-drag'' by analogy to the Coulomb drag effect between two layers of
electrons; this effect is described by the same diagram\cite{Kam95}). 
Third, spurious ultraviolet
divergences occur in the diagrammatic approach to thermal conductivity; these
can be understood and evaluated within the framework of 
divergent series theory\cite{Hardy49}
(their origin is due to illegal series expansions in the derivation of the
perturbation theory).
\setlength{\tabcolsep}{0.2cm}
\renewcommand{\arraystretch}{4}
\begin{table*}
\begin{tabular}{|c|c|c|}
\hline\hline
Dimension&Singlet Term&Triplet Term\\
\hline\hline
3&
$\displaystyle
{105\zeta(5/2)-16\pi^2\zeta(1/2)\over 288\sqrt{2}\pi^{3/2}}
{T^{3/2}\over D^{1/2}}$&
$\displaystyle
{15[(4+7F)\sqrt{1+F}-(4+9F)]\zeta(5/2)
-16\pi^2[2-(2-F)\sqrt{1+F}]\zeta(1/2)\over
96\sqrt{2}\pi^{3/2}F\sqrt{1+F}}
{T^{3/2}\over D^{1/2}}$\\
\hline
2&
$\displaystyle
{T\over 12}\left[\ln{\left({D\kappa_2^2\over T}\right)}
-2\ln{\left({1\over T\tau}\right)}\right]$&
$\displaystyle
-{T\over 2}\left[1-{1\over F}\ln{(1+F)}\right]
\ln{\left({1\over T\tau}\right)}$\\
\hline
1&
&
$\displaystyle
{3\{[(4+3F)\sqrt{1+F}-(4+5F)]-4\pi[(2+F)-2\sqrt{1+F}]\}\zeta(3/2)\over
8\sqrt{2\pi}F}T^{1/2}D^{1/2}$\\
\hline\hline
\end{tabular}
\caption{
Interaction corrections to the electronic thermal conductivity in
one, two and three dimensions in both the singlet and triplet channels.
In the above equations $T$ is temperature, $D$ is the diffusion constant,
$F$ is the effective interaction in the triplet channel, and $\kappa_2$
is the inverse screening length in two dimensions. 
Note that $\hbar=1$ and $k_B=1$
in these calculations. We do not include a result for
the singlet channel in one dimension as it is just one third of that for
the triplet channel with $F$ replaced by the effective interaction parameter
for the singlet channel, $F_s\approx \kappa_3^2a^2\ln{(D\kappa_3^2/T)}$,
where $\kappa_3$ is the inverse screening length in three dimensions, 
and $a$ is the wire width. The inverse screening lengths in two and three
dimensions are given by $\kappa_3^2=8\pi N(0)e^2$ and $\kappa_2=4\pi N(0)e^2$,
where $N(0)$ is the single-spin electronic density of states in the 
appropriate dimension. }
\end{table*}
\renewcommand{\arraystretch}{1}

Before we proceed to the details of our calculation, we present a short
history of the field. We will first consider effects which do not require
interaction, such as weak localization and the Anderson transition. 
Chester and Thellung\cite{Ches60} 
used an exact eigenstates approach to show that the 
Wiedemann-Franz law should hold in a non-interacting disordered system,
independent of the strength of impurity scattering. 
Strinati and Castellani\cite{Str87}
used a Ward identity construction to argue that the Wiedemann-Franz law holds
all the way to the Anderson transition. 
Kearney and Butcher\cite{Kear88} used the exact
eigenstates approach of Chester and Thellung to deduce the weak localization
correction to thermal conductivity. This effect was later seen in the experiment
of Bayot et al\cite{Bayo90}, 
which measured the electrical and thermal magnetoconductance,
which were found to obey the Wiedemann-Franz law. 
Enderby and Barnes\cite{End94} later
pointed out that the Wiedemann-Franz law is not obeyed at the Anderson 
transition -- the previous treatments all use the Sommerfeld expansion, which 
is not valid close to the Anderson transition. However the only effect is to
decrease the constant in the Lorenz number from $\pi^2/3\approx 3.29$ to
$2.17$ for a conductivity exponent $\nu=1$. The overall conclusion for the
non-interacting effects is that there is no extra information in the thermal
conductivity that is not present in the electrical conductivity. In the weak
localization regime the Wiedemann-Franz law holds; close to the Anderson
transition it is only modified by a simple numerical factor, so all critical
properties are correctly predicted by it.

It would therefore seem that the only possibility of interesting behavior lies
with the interaction corrections to thermal conductivity. 
Castellani et al\cite{Cast88} predicted that the 
Wiedemann-Franz law would hold even for the interacting disordered system. 
These authors evaluated the dynamic energy-energy correlation function in the
interacting system using a skeleton graph analysis, and derived the thermal
conductivity from this. Since their method of calculation is somewhat different
from ours, it is not obvious to us why they obtain a different result.
Livanov et al\cite{Liv91} 
later directly calculated the interaction corrections using a
quantum kinetic equation approach, and predicted that the Wiedemann-Franz law
is violated in all dimensions. They appear to have evaluated some interaction
contributions correctly, but to have missed other contributions of the same
order of magnitude; for example, in two dimensions,
they predict a logarithmic increase in 
$\kappa/T$ at low temperature rather than the correct logarithmic decrease.
We believe that the results obtained initially from their diagrammatic formalism 
are correct, and that they have then erroneously thrown away some terms believing 
them to be smaller in magnitude. 
Arfi\cite{Arfi92} 
performed the equivalent calculation in the Matsubara formalism and again
found violation of the Wiedemann-Franz law. However this calculation has errors
relating to all three of the difficulties referred to previously: (a) the
Matsubara form of heat current is used, but the heat current is then erroneously
renormalized by interaction; (b) the ``heat drag'' or Aslamazov-Larkin-like
diagram is omitted; (c) the final results are parametrically larger than those
for electrical conductivity by factors of $1/(T\tau)^2$, where $\tau$ is the
elastic scattering time, which appears to be
due to incorrect treatment of spuriously divergent terms, as will be described
later. The aim of the present work is therefore to resolve the discrepancies
between previous calculations and to present the first consistent calculation
of the interaction correction to thermal conductivity.

The remainder of the paper is organised as follows: in section 2 we derive the
Drude contribution to electronic thermal conductivity; in section 3 we present
an outline of the calculation of the interaction corrections to thermal 
conductivity; finally in section 4 we analyse our results and draw conclusions.
\renewcommand{\arraystretch}{1}

\section{Drude Thermal Conductivity and the Wiedemann-Franz Law}

Before we proceed to the calculation of the interaction corrections, we
demonstrate the simplicity and elegance of the Matsubara approach to
evaluating thermal conductivity by deriving the Drude result. The thermal
conductivity is obtained from the imaginary time heat response kernel,
$Q_{hh}(i\Omega_p)$, by analytic continuation from positive Bose Matsubara
frequencies, $\Omega_p=2\pi Tp$,
\begin{equation}
\label{Qdef}
\kappa T=\lim_{\Omega\rightarrow 0}
\left[{Q_{hh}(i\Omega_p)\over\Omega_p}
\right]_{i\Omega_p\rightarrow\Omega+i0}
\end{equation}
The Feynman diagram for the Drude thermal conductivity is shown in Fig. 1.
\begin{figure}
\includegraphics[width=80mm]{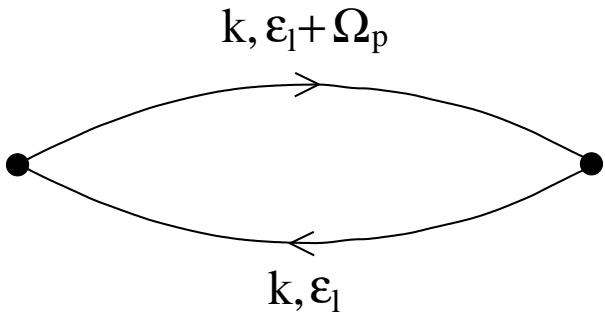}
\caption{Feynman diagram for the Drude contribution to the heat-current
response function. The momentum and Matsubara frequency dependence of the
two electron lines are labelled.}
\end{figure}
The solid lines are disordered electron Green functions
\begin{equation}
\label{Gdef}
G(k,i\varepsilon_l)={1\over i\varepsilon_l-\xi_k+{i\over 2\tau}
\hbox{sgn}(\varepsilon_l)}
\end{equation}
where $\varepsilon_l=2\pi T(l+1/2)$ is a Fermi Matsubara frequency,
$\xi_k=k^2/2m-\mu$ is the electronic excitation spectrum, and $\tau$ is the
elastic scattering time. The black dots represent heat-current vertices, which 
are given by 
\begin{equation}
\label{Jdef}
{\bf j}_h({\bf k},\varepsilon_l,\varepsilon_l+\Omega_p)=
{{\bf k}\over 2m}i(2\varepsilon_l+\Omega_p).
\end{equation}
in the Matsubara frequency representation that we use.
The heat current kernel is then given by
\begin{eqnarray}
\label{Q0}
Q_{hh}^{\alpha\beta} (\mathrm{i} \Omega_p)= 2
T\sum_{\epsilon_l}\sum_k
\frac{k_{\alpha}}{m} \frac{k_{\beta}}{m}
\Big[ \mathrm{i} (\epsilon_l +\Omega_p/2)\Big]^2\times\nonumber\\
G(k, i\epsilon_l)G(k,i\epsilon_l + i\Omega_p)
\end{eqnarray}
Performing the ${\bf k}$-integral, we only obtain a non-zero result if the
frequencies $\epsilon_l+\Omega_p$ and $\epsilon_l$ have opposite sign, which
means that the $\epsilon_l$ is restricted to the range $-\Omega_p<\epsilon_l<0$.
Upon changing the sign of $\epsilon_l$ we obtain
\begin{equation}
\label{Q0a}
Q_{hh}^{\alpha\beta} (\mathrm{i} \Omega_p)
= -4\pi N(0)D\delta_{\alpha\beta}
T\sum_{0<\epsilon_l<\Omega_p}(\epsilon_l-\Omega_p/2)^2,
\end{equation}
where $N(0)$ is the single-spin electronic density of states at the Fermi
surface.
We see that the response function is isotropic, so we drop the spatial indices.
We can then perform the $\epsilon_l$ sum to obtain
\begin{eqnarray}
\label{Q0b}
Q_{hh}&=&-16\pi^3N(0)DT^3\sum_{l=0}^{p-1} (l+1/2-p/2)^2\nonumber\\
&=&-{4\pi^3\over 3}N(0)DT^3(p^3-p),
\end{eqnarray}
and we finally extract $\kappa$ using Eq. (\ref{Qdef}) to get the Drude
result
\begin{equation}
\label{Drude}
\kappa_0={2\pi^2\over 3}N(0)DT={\pi^2n\tau T\over 3}
\end{equation}
where $n$ is the electron number density and we have used the Einstein 
relation $2N(0)D=n\tau/m$. From the corresponding Drude formula for
electrical conductivity, $\sigma_0=ne^2\tau/m$, we see that the Wiedemann-Franz
law is obeyed (note that $\hbar=1$ and $k_B=1$ in our calculations). 

This diagrammatic technique offers a simple proof that the Wiedemann-Franz law
is obeyed for a non-interacting disordered system with arbitrary disorder
strength. For any diagrammatic contribution to electrical conductivity, 
$\sigma$, there is a corresponding contribution to thermal conductivity, 
$\kappa$. Moreover the corresponding expressions differ only in the form of
the current vertices, yielding a factor $k_{\alpha}k_{\beta}e^2/m^2$ for
$\sigma$, and $k_{\alpha}k_{\beta}[i(\epsilon_l+\Omega_p/2)]^2/m^2$ for
$\kappa$. The only essential difference between $\kappa$ and $\sigma$ then
lies in the frequency sums; the ratio of these two sums is independent of
disorder and leads to the Wiedemann-Franz ratio. In particular, the weak
localization correction to thermal conductivity is obtained directly from
the Wiedemann-Franz law. Note that this proof relies
on the Sommerfeld expansion since we are linearising our energy integrals about
the Fermi surface. Interaction effects can violate the Wiedemann-Franz law 
since the presence of an interaction line can alter the Matsubara frequencies 
at the two current vertices.  

\section{The Interaction Corrections}

In this section we calculate the interaction corrections arising from
the singlet and triplet interaction channels \cite{Gril88}
 -- we do not evaluate
Cooperonic contributions since these are expected to be small for a
system with repulsive interactions. In the singlet channel the dominant
contribution arises from small energy and momentum transfers between electrons.
This is dominated by the bare Coulomb interaction, which takes the form
\begin{equation}
\label{V0def}
V_0(q)=\left\{ \begin{array}
{c@{\quad}c}
\displaystyle{4\pi e^2\over q^2}&d=3\\
\displaystyle{2\pi e^2\over q}&d=2\\
\displaystyle e^2\ln{\Big({1\over q^2a^2}\Big)}&d=1 \end{array} \right.
\end{equation}
where $d$ is the dimensionality, and $a$ is a measure of the transverse width
in a quasi-one dimensional wire. The disorder-screened singlet interaction then
takes the form
\begin{equation}
\label{Vsdef}
V_s(q,i\omega_n)={1\over V_0(q)^{-1}+\Pi(q,i\omega_n)},
\end{equation}
where the polarization operator $\Pi(q,i\omega_n)$ is given by
\begin{equation}
\label{Pidef}
\Pi(q,i\omega_n)=2N(0){Dq^2\over Dq^2+|\omega_n|}.
\end{equation}
If any integrals we obtain involving $V_s(q,i\omega_n)$ are convergent at
small momentum, $q$, we can ignore the $V_0(q)^{-1}$
\begin{figure}
\includegraphics[width=90mm]{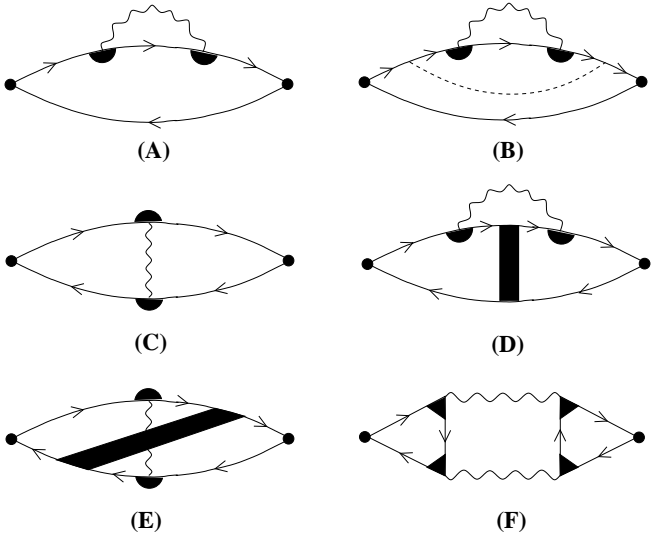}
\caption{Feynman diagrams for the interaction correction to the
heat-current response function. Diagrams A--E are structurally identical to
the diagrams which contribute to electrical conductivity. Diagram F is the
additional ``heat-drag'' diagram which must be included to obtain a
consistent result.}
\end{figure}
 term in 
Eq. (\ref{Vsdef}) in comparison to $\Pi(q,i\omega_n)$; $V_s(q,i\omega_n)$
then takes the universal form,
\begin{equation}
\label{Vuniv}
2N(0)V_s(q,i\omega_n)={Dq^2+|\omega_n|\over Dq^2}.
\end{equation}
In the triplet channel the dominant contribution arises from momentum transfers
of the order of $2k_F$, and the unsceened triplet interaction can be treated
as a constant. The disorder screened triplet interaction then takes the form
\begin{equation}
\label{Vtdef}
2N(0)V_t(q,i\omega_n)=F{Dq^2+|\omega_n|\over
(F+1)Dq^2+|\omega_n|},
\end{equation}
where $F=2N(0)V_t^0$ and $V_t^0$ is the bare interaction in the 
triplet channel\cite{Fdef}.
Note that $F$ includes Fermi liquid corrections in the absence of disorder (it
is only unscreened with respect to the disorder). $F$ may be determined from
the measured paramagnetic spin susceptibility in the experimental system of
interest
\begin{equation}
\label{chidef}
\chi={2N(0)\over 1+F}
\end{equation}
If we first calculate the contribution from the singlet channel, we can then
obtain the corresponding result for the triplet channel by replacing $V_s$
by $V_t$ and multiplying by $3$ -- the extra factor of $3$ arises because 
there are $3$ times as many modes in the spin-one channel than in the spin-zero
channel.

Applying standard perturbation theory we find that the leading order interaction
corrections are given by the Feynman diagrams shown in Fig. 2.
These diagrams are identical to those considered by Altshuler and Aronov for
the electrical conductivity\cite{AA79}, with the exception of diagram (F) which we call
the ``heat-drag'' term. The corresponding diagram for electrical conductivity
is a factor $(T/E_F)^2$ smaller than the other diagrams due to cancellation
of electron and hole charge currents; the heat current has opposite electron-hole
parity and thus electron and hole heat currents reinforce each other.
Consequently diagram (F) is of the same size as the other diagrams in the case
of thermal conductivity. The use of the Matsubara heat current vertex defined
in Eq. (\ref{Jdef}) ensures that no interaction renormalization of vertices
is needed, greatly simplifying the calculation with respect to other choices
of heat current operator (obviously the final results must be independent of
this choice). 

Details of the calculation of the contributions from diagrams A--F to the
heat-current response function, $Q_{hh}(i\Omega)$, are presented in the 
appendix (note that in the following we suppress the subscripts on
Matsubara frequencies $\Omega_p=2\pi Tp$ and $\omega_n=2\pi Tn$ for
convenience). It is shown there that $Q_{hh}(0)$ vanishes, as it must for
internal consistency, and that $\kappa T$ may be written as the limit of
the expression 
\begin{widetext}
\begin{eqnarray}
\label{kappaT}
&&+4N(0)D{T\over\Omega}\sum_{0<\omega\le\Omega}\sum_q\omega^3
\left[1-{8\over 3d}{Dq^2\over (Dq^2+\omega)}\right]
{V(q,\omega)\over (Dq^2+\omega)^2}\nonumber\\
&&-{16\over 3d}\pi^2T^2N(0)D\left[T\sum_{\omega>0}+
{T\over\Omega}\sum_{0<\omega\le\Omega}\omega\right]\sum_q
{Dq^2V(q,\omega)\over (Dq^2+\omega)^3}
\end{eqnarray}
\begin{eqnarray}
&&+{16\over d}N(0)^2DT\sum_{\omega>0}\sum_q\omega^3
\left[{2Dq^2V(q,\omega)\over (Dq^2+\omega)}
+\omega{\partial V(q,\omega)\over\partial\omega}\right]
{Dq^2V(q,\omega)\over (Dq^2+\omega)^4}\nonumber
\end{eqnarray}
\end{widetext}
as $\Omega$ tends to zero. As explained in the appendix, this statement is
to be interpeted in the sense of first continuing $i\Omega_p$ to real
frequencies, and then taking the limit of real frequency going to zero.
The second line of Eq. (\ref{kappaT}) is just the interaction contribution
to the electrical conductivity multiplied by the factor $\pi^2T^2/3e^2$.
In other words, this is the contribution to $\kappa T$ predicted by the
Wiedemann-Franz law. The other two terms thus lead to violation of the
Wiedemann-Franz law if they yield a non-zero result (as we indeed find that
they do). The expressions in Eq. (\ref{kappaT}) may now be evaluated in
one, two and three dimensions for both the singlet and triplet interactions
to yield the results listed in Table 1. Note that all three terms
in Eq. (\ref{kappaT}) give results having the same parametric form but with
different constant prefactors; we must therefore evaluate all of them to
get a correct final result. This statement is not quite true in two dimensions
because of the presence of logarithmic terms, and we should keep the most
singular terms in this case.

\section{Conclusions}
We have calculated the interaction corrections to thermal conductivity in
the diffusive regime of a disordered conductor. Our main result is that the
Wiedemann-Franz law is violated in all dimensions; the predicted interaction
results have the expected parametric dependences, but different numerical
coefficients. For example, in the singlet channel in two dimensions, the
actual logarithmic correction is half that predicted by the Wiedemann-Franz
law. Unfortunately the experimental work of Bayot et al\cite{Bayo90}
is the only work we know of on thermal conductivity in disordered
conductors, and this only isolates a weak localization contribution. We hope
that further experiments will be performed in this area, and that the 
interaction correction be observed as well as the weak localization correction.
In particular the two-dimensional system would seem to be a promising one
to investigate. Bayot et al's\cite{Bayo90} disordered graphitic system
showed weak localization effects whose magnitude was roughly 20\% of the
Drude term in the electronic thermal conductivity at a temperature of $2.9K$.
If it were possible to cleanly extract the phonon term, one could 
look at the electronic term as a function of both temperature and magnetic 
field, and distinguish weak localization and interaction effects. This would
then experimentally settle the question of whether the Wiedemann-Franz law
is violated in an interacting disordered system. Another sensible quantity
to investigate experimentally would be the thermal Hall conductivity, which
arises solely from electrons (although there can be phonon drag effects).
In future work we intend to derive the interaction corrections to this
quantity.

\begin{acknowledgments}

We thank A.M. Finkel'stein, I.V. Lerner and I.V. Yurkevich for
helpful discussions. We acknowledge support from the UK EPSRC.

\end{acknowledgments}

\appendix*
\section{Diagrammatic Contributions}

In this appendix we list the complete set of diagrammatic contributions for
archival purposes. This should allow interested researchers to reproduce our
results in detail. The total contribution to the heat-current response
function, $Q_{hh}(i\Omega)$, from all the diagrams in Fig. 2 is given by 
\begin{widetext}
\begin{equation}
\begin{array}{l@{\qquad\qquad}l}
-\displaystyle 8\pi N(0)D\delta_{\alpha\beta}T\sum_{\omega>0}
T\sum_{0<\epsilon<\omega}
\sum_q \left[\epsilon+{\Omega\over 2}\right]^2
{V(q,\omega)\over (Dq^2+\omega)^2}
&\hbox{(A1)}\\
-\displaystyle 8\pi N(0)D\delta_{\alpha\beta}T\sum_{\omega>\Omega}
T\sum_{0<\epsilon<\omega-\Omega}\sum_q
\left[\epsilon+{\Omega\over 2}\right]^2
{V(q,\omega)\over (Dq^2+\omega)^2}
&\hbox{(A2)}\\
\displaystyle +8\pi N(0)D\delta_{\alpha\beta}T\sum_{0<\omega\le\Omega}
T\sum_{0<\epsilon<\omega}\sum_q\left[\epsilon-{\Omega\over 2}\right]^2
{V(q,\omega_n)\over (Dq^2+\omega)^2}
&\hbox{(A3)+(B)}\\
\displaystyle +8\pi N(0)D\delta_{\alpha\beta}T\sum_{\omega>\Omega}
T\sum_{0<\epsilon<\Omega}\sum_q\left[\epsilon-{\Omega\over 2}\right]^2
{V(q,\omega)\over (Dq^2+\omega)^2}
&\hbox{(A3)+(B)}\\
\displaystyle +16\pi N(0)D\delta_{\alpha\beta}T\sum_{\omega>\Omega}
T\sum_{0<\epsilon<\omega-\Omega}\sum_q
\left[\epsilon+{\Omega\over 2}\right]
\left[\epsilon+{\Omega\over 2}-\omega\right]
{V(q,\omega)\over (Dq^2+\omega)^2}
&\hbox{(C1)+(C2)}
\end{array}
\end{equation}
\begin{displaymath}
\begin{array}{l@{\qquad\qquad}l}
\displaystyle +32\pi N(0)DT\sum_{\omega>0}T\sum_{0<\epsilon<\omega}\sum_q
\left[\epsilon+{\Omega\over 2}\right]^2
{Dq_{\alpha}q_{\beta}V(q,\omega)\over
(Dq^2+\omega)^2(Dq^2+\omega+\Omega)}
&\hbox{(D1)}\\
\displaystyle +32\pi N(0)DT\sum_{\omega>\Omega}
T\sum_{0<\epsilon<\omega-\Omega}\sum_q
\left[\epsilon+{\Omega\over 2}\right]^2
{Dq_{\alpha}q_{\beta}V(q,\omega)\over
(Dq^2+\omega)^2(Dq^2+\omega-\Omega)}
&\hbox{(D2)}\\
\displaystyle -32\pi N(0)DT\sum_{\omega>\Omega}
T\sum_{0<\epsilon<\omega-\Omega}\sum_q
\left[\epsilon+{\Omega\over 2}\right]
\left[\epsilon+{\Omega\over 2}-\omega\right]
{Dq_{\alpha}q_{\beta}V(q,\omega)\over
(Dq^2+\omega)^2(Dq^2+\omega+\Omega)}
&\hbox{(E1)}\\
\displaystyle -32\pi N(0)DT\sum_{\omega>\Omega}
T\sum_{0<\epsilon<\omega-\Omega}\sum_q
\left[\epsilon+{\Omega\over 2}\right]
\left[\epsilon+{\Omega\over 2}-\omega\right]
{Dq_{\alpha}q_{\beta}V(q,\omega)\over
(Dq^2+\omega)^2(Dq^2+\omega-\Omega)}
&\hbox{(E2)}\\
\displaystyle +16N(0)^2DT\sum_{\omega>0}\sum_q\omega^2(\omega+\Omega)^2
{Dq_{\alpha}q_{\beta}V(q,\omega)V(q,\omega+\Omega)\over
(Dq^2+\omega)^2(Dq^2+\omega+\Omega)^2}
&\hbox{(F)}
\end{array}
\end{displaymath}
\end{widetext}

In the above expression we have listed the contributions according to which
diagram they are obtained from; the additional numerical label following the
letter refer to different sign configurations of Matsubara frequencies which
are possible within the same diagram. For example, in diagram (A), there are
three possible sign configurations whose contributions we denote by $(A1)$,
$(A2)$ and $(A3)$. To proceed further we next perform the sums over the
Fermi Matsubara frequency, $\epsilon$. 
After some simplification we then obtain the following result for
$Q_{hh}(i\Omega)$:
\begin{widetext}
\begin{equation}
\begin{array}{l@{\qquad\qquad}l}
\displaystyle -4N(0)D\delta_{\alpha\beta}T\sum_{\omega>\Omega}\sum_q
\omega^3{V(q,\omega)\over (Dq^2+\omega)^2}
-4N(0)D\delta_{\alpha\beta}T\sum_{0<\omega\le\Omega}\sum_q
\omega^2\Omega{V(q,\omega)\over (Dq^2+\omega)^2}
&\hbox{(ABC)}\\
\displaystyle +16N(0)DT\sum_{\omega>\Omega}\sum_q
\omega^2(\omega-\Omega){Dq_{\alpha}q_{\beta}V(q,\omega)\over
(Dq^2+\omega)[(Dq^2+\omega)^2-\Omega^2]}
&\hbox{(DE)}\\
\displaystyle +{4\over 3}N(0)DT\sum_{\omega>\Omega}\sum_q
\big[\Omega^3+12\omega^2\Omega-4\pi^2T^2\Omega\big]
{Dq_{\alpha}q_{\beta}V(q,\omega)\over
(Dq^2+\omega)^2(Dq^2+\omega+\Omega)}
&\hbox{(DE)}\\
\displaystyle +{4\over 3}N(0)DT\sum_{0<\omega\le\Omega}\sum_q
\big[4\omega^3+6\omega^2\Omega+3\omega\Omega^2-4\pi^2T^2\omega\big]
{Dq_{\alpha}q_{\beta}V(q,\omega)\over
(Dq^2+\omega)^2(Dq^2+\omega+\Omega)}
&\hbox{(DE)}\\
\displaystyle +16N(0)^2DT\sum_{\omega>0}\sum_q\omega^2(\omega+\Omega)^2
{Dq_{\alpha}q_{\beta}V(q,\omega)V(q,\omega+\Omega)\over
(Dq^2+\omega)^2(Dq^2+\omega+\Omega)^2}
&\hbox{(F)}
\end{array}
\end{equation}
\end{widetext}
Note that the contributions from diagrams A, B and C do not cancel each other
as they do in the corresponding electrical conductivity calculation.
At this point we can check that we have chosen a consistent set of diagrams
to evaluate by setting the external frequency, $\Omega=0$, and checking that
$Q_{hh}(0)=0$. This is a very powerful test that should not be omitted -- it
is very dangerous to merely calculate $[Q_{hh}(i\Omega)-Q_{hh}(0)]/\Omega$,
especially when a large number of diagrams are involved. One always runs the
risk of missing important physical processes; in fact we only became aware of
the presence of the heat-drag term of diagram (F) when this check failed in
its absence. Setting $\Omega=0$ here yields
\begin{widetext}
\begin{eqnarray}
\label{Qint0}
Q_{hh}(0)&=&4N(0)DT\sum_{\omega>0}\sum_q
\left[-\delta_{\alpha\beta}+{4Dq_{\alpha}q_{\beta}\over (Dq^2+\omega)}
+{4\omega Dq_{\alpha}q_{\beta}N(0)V(q,\omega)\over (Dq^2+\omega)^2}\right]
{\omega^3V(q,\omega)\over (Dq^2+\omega)^2}
\nonumber\\
&=&-2T\sum_{\omega>0}\sum_q\omega^2
{\partial^2\over\partial q_{\alpha}\partial q_{\beta}}
\ln{\left[V_0(q)^{-1}+\Pi(q,\Omega)\right]}=0
\end{eqnarray}
\end{widetext}
where in the last step we have used the divergence theorem to convert the 
$q$-integral into a surface integral with its bounding surface at infinity.
Note that this check not only gives us confidence that all relevant diagrams
have been included, but also that each diagram has been given the correct
combinatorial factors. In the above derivation
we have treated $V_0(q)^{-1}$ as though it had no $q$-dependence. This is
justified as the terms ignored are smaller in magnitude, and would be
cancelled by higher order diagrams. In addition to the above proof that
$Q_{hh}(0)=0$, we have also directly evaluated $Q_{hh}(0)$ from
Eq. (\ref{Qint0}) in one, two, and three dimensions, for both singlet and
triplet potentials, and found it to equal to zero. 

We can then find $\kappa T$ as the limit of $Q_{hh}(\Omega_p)/\Omega_p$
as $\Omega_p$ tends to zero. Now of course this statement does not make
sense since $\Omega_p$ is discrete -- we should first analytically continue
to real frequencies and then take the limit. However it turns out to be
possible to avoid performing this analytic continuation, and instead
to continue
manipulating discrete sums. We can then extract the term that is proportional
to $\Omega_p$ and discard terms proportional to higher powers of $\Omega_p$.
This approach is legitimate provided that any operation we perform on
$\Omega_p$ would carry over unchanged to the same operation on the
corresponding real
frequency after analytic continuation. The advantage of this method is
simply convenience in calculation -- no illegal operations occur, as we
have checked by performing the analytic continuation first and then taking
the zero-frequency limit. This procedure leads to the expression for
$\kappa T$ as the zero-frequency limit of
\begin{widetext}
\begin{eqnarray}
\label{kappaT1}
&&+4N(0)D{T\over\Omega}\sum_{0<\omega\le\Omega}\sum_q\omega^3
\left[1-{8\over 3d}{Dq^2\over (Dq^2+\omega)}\right]
{V(q,\omega)\over (Dq^2+\omega)^2}\nonumber\\
&&-{16\over 3d}\pi^2T^2N(0)D\left[T\sum_{\omega>0}+
{T\over\Omega}\sum_{0<\omega\le\Omega}\omega\right]\sum_q
{Dq^2V(q,\omega)\over (Dq^2+\omega)^3}\\
&&+{16\over d}N(0)^2DT\sum_{\omega>0}\sum_q\omega^3
\left[{2Dq^2V(q,\omega)\over (Dq^2+\omega)}
+\omega{\partial V(q,\omega)\over\partial\omega}\right]
{Dq^2V(q,\omega)\over (Dq^2+\omega)^4}\nonumber
\end{eqnarray}
\end{widetext}
We can then perform the $q$-integral using the standard identification
\begin{equation}
\sum_q=\int {d^dq\over (2\pi)^d}
\end{equation}
and the formulae for the $V(q,i\omega)$ given in Eq. (\ref{Vsdef}) and
Eq. (\ref{Vtdef}). After performing this $q$-integral we end up with terms
which are infinite sums over powers of $\omega$. Provided that the sums are
ultraviolet divergent, we can identify them as zeta functions via
\begin{equation}
T\sum_{\omega>0}\omega^k=
T(2\pi T)^k\sum_{n>0}n^k=
T(2\pi T)^k\zeta(-k).
\end{equation}
Any sum that is infrared divergent must be cut off correctly since such
a divergence is physical -- such a situation occurs in two-dimensions where
we obtain a logarithmic sum which is cut off at $\omega\sim T$ at low
frequency and $\omega\sim 1/\tau$ at high frequency. We have used this 
particular divergent series trick as it allows a very direct evaluation 
of results. As a check of its legality we have recalculated the various
terms using standard analytic continuation methods and obtained the same
results -- albeit after a lot more algebraic manipulation.



\begin{thebibliography}{16}
\bibitem{Fdef} Note that there are two different definitions of parameter 
$F$ in the literature; we follow that of Ref. [2] rather than that of Ref. [1]
in this paper. The $F$ of Ref. [1] is equal to $-2F/(F+1)$ in our language.
\end{thebibliography}
\end{document}